\documentclass[11pt]{article}
\usepackage{graphicx}
\usepackage{caption}
\usepackage{subcaption}
\usepackage{tabularx}
\usepackage{layouts}
\usepackage{multicol}
\usepackage{tikz}
\usepackage{color}
\usepackage{listings}
\usepackage[colorlinks = true,
            linkcolor = blue,
            urlcolor  = blue,
            citecolor = blue,
            anchorcolor = blue]{hyperref}
\usepackage[left=20mm,right=20mm,top=33.95mm,bottom=33.95mm]{geometry}
\usepackage[english]{babel}
\usepackage[utf8x]{inputenc}
\usepackage[T1]{fontenc}
\usepackage{amsmath}
\usepackage[colorinlistoftodos]{todonotes}
\usepackage{authblk}
\usepackage{float} 

\title{
		\usefont{OT1}{bch}{b}{n}
		\Large Vit-GAN: Image-to-image Translation with Vision Transformes \\
		and\\
		Conditional GANS
}

\usepackage{authblk}

\author[1]{Yi\u{g}it G\"und\"u\c{c}}

\affil[1]{\small{ygunduc@gmail.com}}

\begin{document}
\selectlanguage{english}

\maketitle

\begin{abstract}
  
  In this paper, we have developed a general-purpose architecture, Vit-Gan, capable of performing most of the image-to-image translation tasks from semantic image segmentation to single image depth perception. This paper is a follow-up paper, an extension of generator based model~\cite{Gunduc:2021} in which the obtained results were very promising. This opened the possibility of further improvements with adversarial architecture.  We used a unique vision transformers-based generator architecture and Conditional GANs(cGANs) with a Markovian Discriminator (PatchGAN) (\href{https://github.com/YigitGunduc/vit-gan}{https://github.com/YigitGunduc/vit-gan}). In the present work, we use images as conditioning arguments. It is observed that the obtained results are more realistic than the commonly used architectures.  
  
\end{abstract}

\section{Introduction}

Most of the problems we face in computer vision can directly or indirectly benefit from image to image translation. The image-to-image translation problem is mapping an input image to a corresponding output image. Image-to-Image translation has a wide range of applications, such as object transfiguration, season transfer, image segmentation, photo enhancement, or in general, where an image is needed to be mapped onto another. Most of the traditional translation methods are only applicable to a single domain or trained for a special task~\cite{Efros:2001, Hertzmann:2001, Fergus:2006, Buades:2005, Chen:2009, Shih:2013, Laffont:2014, Long:2015, Eigen:2015, Xie:2015, Zhang:2016}. In this work, we propose a general-purpose architecture that is capable of performing most image-to-image translation tasks, from semantic image segmentation to single image depth precision.

\setlength{\parskip}{1em}

Optimization and loss metric is one of the biggest problems in deep learning.  The same problem also exists in the Image-to-Image translation.  In our experiment, we see that the models, which are optimized with the Mean Absolute Error or Mean Square Error functions tend to generate blurry results. This problem is related with the optimization procedure which averages over the values.  Hence such a model is unable to perform well on complex tasks. GANs~\cite{Goodfellow:2014} are the perfect fit for sharp and realistic images as output. In a GAN model, the main idea is; the discriminator learns to classify if the output image is real or fake, and the generator tries to fool it. During simultaneous training, both components play a min-max game and minimize their losses. Generators in GAN models do not learn how to generate an image but rather the loss function itself the discriminator. GANs classify images as real or fake. That makes them powerful tools for the generation of images in as close resemblance as the real ones. 

\setlength{\parskip}{1em}

The proposed solution applies to nearly all image-to-image translation tasks and problems with the help of a relatively new concept of transformers, specifically vision transformers. In this work, we have explored how vision transformers can be used for image-to-image translation. In the current work, we used CNNs alongside patches, patch encoder, and transformers. In this paper, we introduced a unique vision transformers-based generator and Conditional GANs(cGANs)~\cite{Mirza:2014} with a Markovian discriminator (PatchGAN)~\cite{Li:2016}. This discriminator has been applied to similar problems and shown promising results~\cite{pix2pix2017}. We also used the PatchGAN discriminator, which is applied to reinforce our generator to generate sharp and accurate images on difficult cases. 

\section{Background}

In the literature, a variety of algorithms  proposed for image-to-image translation~\cite{Pang:2021} (and references therein). The majority of the methods were based on  CNNs and employed encoder-decoder networks~\cite{Efros:2001} for image-to-image translation. In an autoencoder architecture, the input image passes through the encoder. The encoder layers downsample the input until a bottleneck layer. In a reversed process, starting from the bottleneck point, the decoder upsamples the image until the original shape is restored. The problem with encoder-decoder architectures is that low-level information is lost while the input images pass through the bottleneck.  The information loss may have negative effects while constructing the output image. Some successful methods also utilized U-Net~\cite{Ronneberger:2015} like approaches to overcome this problem. U-Net adds skip connections between layers. if one wants to formulate skip connections between the $i^{\rm th}$ and the $(n − i)^{\rm th}$ layers, where $n$ is the layer number,  one need to create a link in between which acts as attention.

\setlength{\parskip}{1em}

For image-to-image translation, Generative Adversarial Network~\cite{Heusel:2017} (GAN) based approaches also have been used.  A GAN is a generative model, which generates new instances of data used for training. A GAN~\cite{Heusel:2017} consist of two-part a generator and a discriminator. The discriminator is used to lead the generator the generate shapes and accurate images. These networks play a min-max game against each other to minimize their losses. Through the competition, GANs can generate realistic output images from random noise input. Conditional generative adversarial network~\cite{Mirza:2014} is a GAN architecture that enables the generation of output images according to introduced conditions. The condition can be a class label or any other descriptive information on the dataset. 
 
\setlength{\parskip}{1em}

Vaswani et al. proposed Transformer architectures for machine translation~\cite{Vaswani:2017}. Since then, the Transformers became the state of the art in most of the NLP tasks. The Transformer models are often pre-trained on large text corpora and then fine-tuned for specific tasks~\cite{Devlin:2019, Brown:2020, Radford:2018}. Transformers, with their unique self-attention mechanisms and MLPs, do not suffer from memory shortages.  Comparing with the Long Short-Term Memory Networks (LSTMs), which cannot recall tokens after a certain amount of time, transformers show their advantages. There have been attempts to utilize transformers like self-attention architectures with CNNs for computer vision tasks~\cite{Hu:2018, Locatello:2020, Chen:2020a, Li:2019, Sun:2019}. 

\setlength{\parskip}{1em}
  
Image GPT(iGPT)~\cite{Chen:2020b} is another notable architecture that takes pixel sequences as inputs and tries to generate output. Unlike vision Transformers, where they take all the images and split them into patches, iGPT treats pixels just like tokens of words. iGPT takes some pixels and outputs some pixels similar to how a text generation model would work.

\setlength{\parskip}{1em}

Vision transformers (Vit)~\cite{Buades:2005} is a transformer architecture that only uses transformer encoders and a custom patch embedding to work with images. Vision Transformers are used for image classification and reached an accuracy of 88.55\% on ImageNet. Vit works by splitting an image into patches, flattens those patches, produces lower-dimensional linear embeddings from the flattened patches, adding the positional embeddings, and then feeding those sequences as an input to a transformer encoder. Vit models are usually pre-trained on a large dataset and then finetuned on smaller datasets for image classification.

\section{Method}

\subsection{Overview}

The most significant difference between our approach and the other image-to-image translation methods is its unique generator: A hybrid architecture that utilizes vision transformers alongside convolutional layers to convert images~\cite{Gunduc:2021}.  Unlike prior methods, the proposed method utilizes vision transformers for image understanding and uses PatchGAN~\cite{Li:2016} to ensure generated results are clear, accurate, and sharp. In the present work, we use images as conditioning arguments. The generator uses the condition input image to make a selective prediction.   The discriminator takes the output of the generator as input. The discriminator output and L1 loss are used in conjunction to compute the loss. Lambda, the coefficient which accompanied the L1 loss, is chosen as $100$. This coefficient can be tuned if the model generates visual artifacts. This situation never happened in our experiments. For complex tasks, a cGAN alongside L1 loss gives better image quality~\cite{pix2pix2017}. Finally, the optimizer uses the total loss for applying gradients to the weights. Please see, \href{https://github.com/YigitGunduc/vit-gan}{https://github.com/YigitGunduc/vit-gan} for details of the model.

\subsection{Patches and Patch Encoding}

Our generator uses Vit-like patches and patch embedding to understand the image. The model takes inputs, splits them into patches. The number of patches depends on the patch size and image size ( ${\rm image\; height} \times {\rm image\; width}$). The formula $(Image height/patch size)**2$ (assuming image height and width are the same)  is used to calculate the number of patches. Then, all patches are flattened and reshaped into  $1-D$ sequences of patches. We use a trainable embedding layer to embed patches to latent space vectors and add a learnable $1-D$ position embedding to the patches. This last embedding preserves the positions of the patches in the image. 

\subsection{Transformer Layers}

The Generator consists of $N$  identical stacked layers~\cite{Gunduc:2021}. Each stacked-layer consists of a multi-head self-attention (MHA) trailed by a feed-forward linear layer.  MHA contains $H$ heads where each one of the identical heads computes independent scaled dot-product attention. Therefore the model can attend different places of the embedding vector. Then transformation is applied to concatenate the attention results obtained from heads before further processing in the model.

\setlength{\parskip}{1em}

\[ {\rm MultiHead}(Q, K, V ) = {\rm Concat}({\rm head}_1,\dots, {\rm head}_h)W^O \]

\setlength{\parskip}{1em}

The utilized attention is the same as used in the original paper~\cite{Vaswani:2017}.

The  proposed and formula is as follows:

\setlength{\parskip}{1em}

\[{\rm Attention}(Q, K, V ) = {\rm Softmax}\left (\frac{ Q K^T }{\sqrt{ d_k}} \right )V \]

 where, $Q$, $K$, $V$ are the query, key, and the value respectively.

\setlength{\parskip}{1em}

After the dot-product attention is applied we scale the outputs since the dot-product attention tends to output large values. After the normalization process,  finalize the outputs with a feed-forward linear layer.

\subsection{Architecture}

\subsubsection{Generator}

The generator section of the architecture is taken from a previous paper on image-to-image translation. The success of the generator on  image translation has already been
proven.~\cite{Gunduc:2021}. Inputs of the generator are introduced to the pach encoders.  After images are split into patches and passed through the patch encoder, the resulting vectors are input to the transformer encoders.  $N$ of the transformer encoders are stacked on top of each other, which are used to have an understanding of the context. In this image-to-image translation task, this context is the image. The transformer encoder's output is passed to a residual block. The residual block consists of two convolutional layers with skip connections in between. ReLU and batch normalization are applied in the residual blocks. Residual blocks are followed by an upsampling block. Transpose convolutions are trailed by LeakyReLU and batch normalization in the upsampling block.

\setlength{\parskip}{1em}

The experiments with various generator architectures are carried out and it is concluded that using a combination of Residual blocks and transpose convolutions gives the best output~\cite{Gunduc:2021}.

\subsubsection{Discriminator (PatchGAN)}

A carefully chosen loss function plays the utmost importance in the performance of the model.  pix2pix~\cite{pix2pix2017} paper has experimented on this topic and shown that L1 loss with a cGAN outperforms better than the other methods. Hence, we applied the same optimization strategy in our model with L1 loss alongside a cGAN.

\setlength{\parskip}{1em}

If the discriminator is not conditional, the model tends to generate results that fool the discriminator but,  the output images resemble any one of the training data set. In other words, the model treats the input image as a random noise like most of the GANs algorithms do. This situation is not acceptable for image-to-image translation tasks. The output must be built upon the input and share the same low-level features.

\setlength{\parskip}{1em}

PatchGAN is a discriminator for GANs which penalizes local image patches rather than the whole image. Discriminator tries to classify $N \times N$ patches as real or fake.  Classifying patches rather than the whole image enables  PatchGAN to run and train much faster.  In our experimentation with the model, we also concluded that PatchGAN outperforms traditional discriminators. Hence PatchGAN is selected as the discriminator for our model.

\section{Experiments} 

We tested our method on a different task to see how well it performed. We used image segmentation, depth prediction from a single image, architectural labels to photos.  After training, the model performed excellently. We concluded that such a performance is related to the data set.  

\subsection{Datasets} 

We have tested our method on three different datasets for three different Tasks. Cityscapes~\cite{Cordts:2016,Cordts:2015}, Facades~\cite{Tylecek:2013}, and RGB-D~\cite{Sturm:2012} datasets are used to test, models prediction ability of semantic segmentation,  convert architectural labels to facades and  depth maping from a single image.

\begin{figure}[th!]
  \centering
  \begin{subfigure}[b]{\textwidth}
    \centering
    \includegraphics[width=\textwidth,height=8cm]{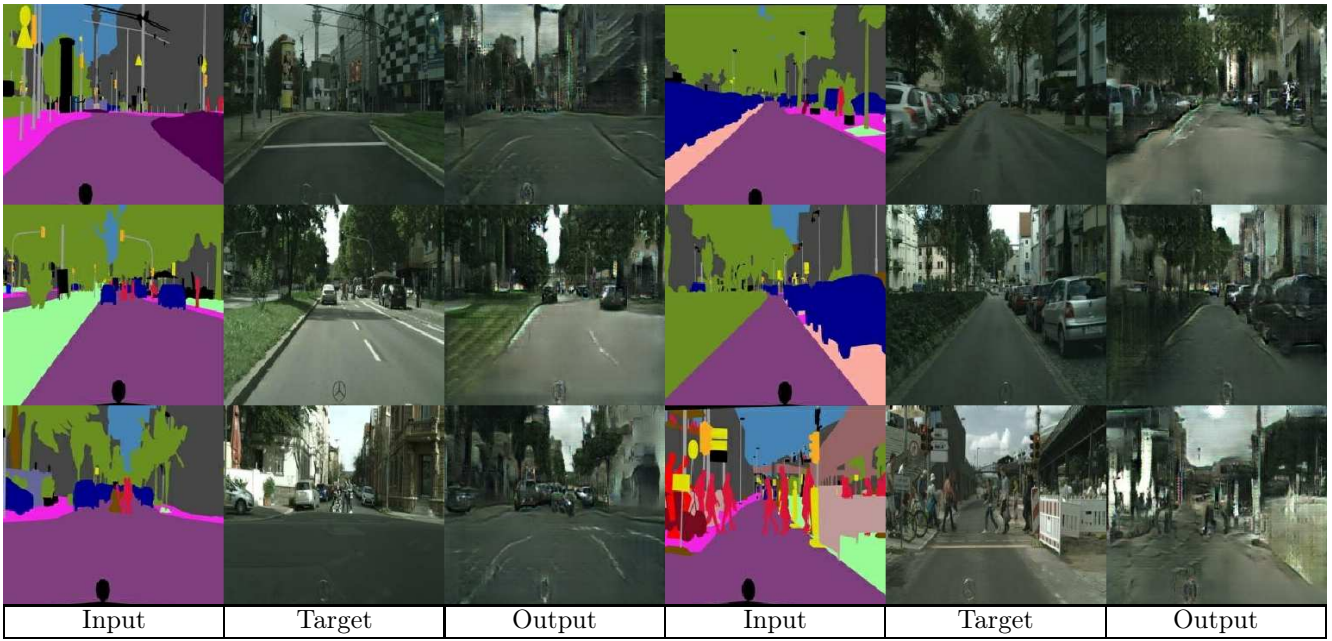}
    \vspace*{-7mm}
    \caption*{\begin{tabularx}{\textwidth}{
          | >{\centering\arraybackslash}X
          | >{\centering\arraybackslash}X
          | >{\centering\arraybackslash}X
          | >{\centering\arraybackslash}X
          | >{\centering\arraybackslash}X
          | >{\centering\arraybackslash}X | }
        \hline  
        Input & Target & Output & Input & Target & Output\\
        \hline \end{tabularx}}
  \end{subfigure}
  \caption{Generation of a photo of the scene when a segmentation map has been inputted }
  \label{fig:Combine_c}
\end{figure}

\textbf{Cityscapes}~\cite{Cordts:2016,Cordts:2015} is a large-scale database that is explicitly built for semantic segmentation. The database contains urban street scenes which provide semantic instances of 30 classes grouped into 8 categories (flat surfaces, humans, vehicles, constructions, objects, nature, sky, and void). The dataset consists of around 5000 annotated images. Images from 50 cities in Germany with different lighting and weather conditions constitute the dataset. We trained our model by using this dataset for both ways, to generate semantic segmentation maps (Figure~\ref{fig:Combine_s}) when the photo is inputted and generate a photo of the scene when a segmentation map has been inputted (Figure~\ref{fig:Combine_c}).

\setlength{\parskip}{1em}

\textbf{Facades}~\cite{Tylecek:2013} is a database of facade images with its corresponding labels of machine perception. The dataset includes 606 annotated rectified images from diverse parts of the world with distinct architectural styles of facades obtained from different sources.  

\begin{figure}[th!]
  \centering
  \begin{subfigure}[b]{\textwidth}
    \centering
    \includegraphics[width=\textwidth,height=8cm]{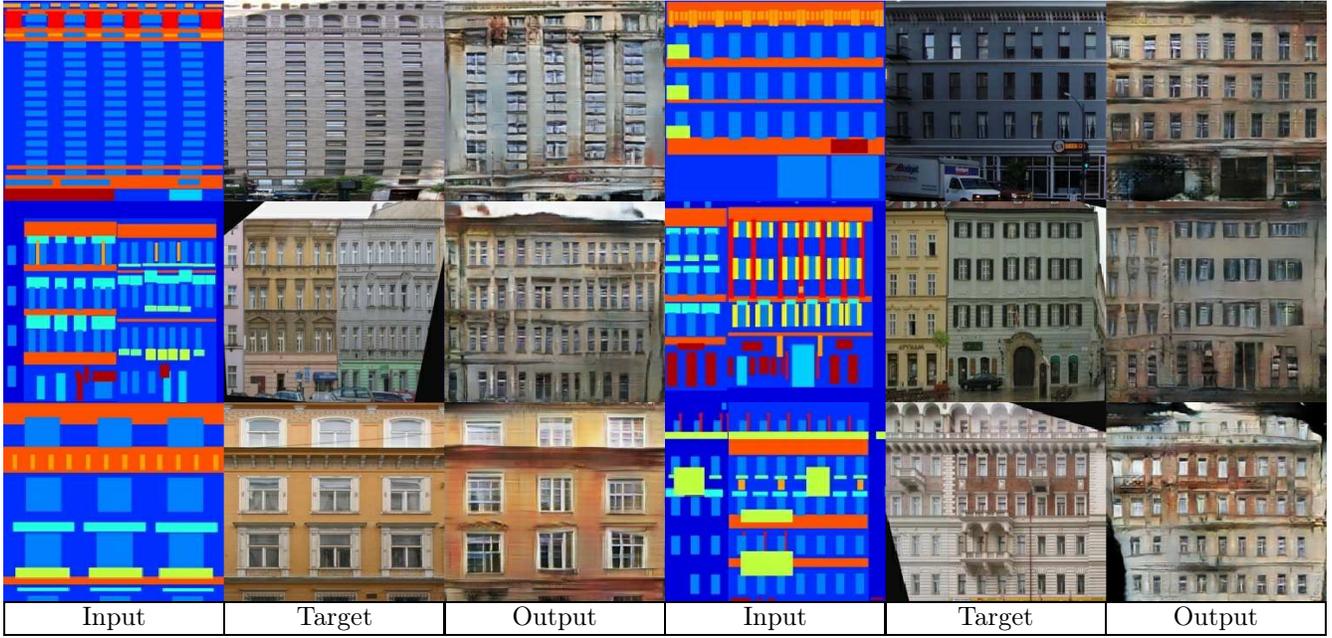}
    \vspace*{-7mm}
    \caption*{\begin{tabularx}{\textwidth}{
          | >{\centering\arraybackslash}X
          | >{\centering\arraybackslash}X
          | >{\centering\arraybackslash}X
          | >{\centering\arraybackslash}X
          | >{\centering\arraybackslash}X
          | >{\centering\arraybackslash}X | }
        \hline  
        Input & Target & Output & Input & Target & Output\\
        \hline \end{tabularx}}
  \end{subfigure}
  \caption{Facets dataset: Converting architectural labels to facades }
  \label{fig:Combine_f}
\end{figure}

Our model has been trained on the Facades dataset to convert architectural labels to facades (Figure~\ref{fig:Combine_f}).

\setlength{\parskip}{1em}

\textbf{RGB-D}~\cite{Sturm:2012} dataset consists of various indoor and outdoor images with associated depth maps. Indoor and outdoor samples are taken from offices, rooms, dormitories, exhibition centers, streets, the road from Yonsei University and Ewha University. The model is trained on the RGB-D dataset to experiment with the depth prediction precision from a single image. The model successfully generates depth maps from a single raw input image (Figure~\ref{fig:RGB-D_Depth}).

\subsection{Experiment on Custom Disciriminator}

\begin{figure}
 \begin{center}
   \includegraphics[width=0.9\textwidth]{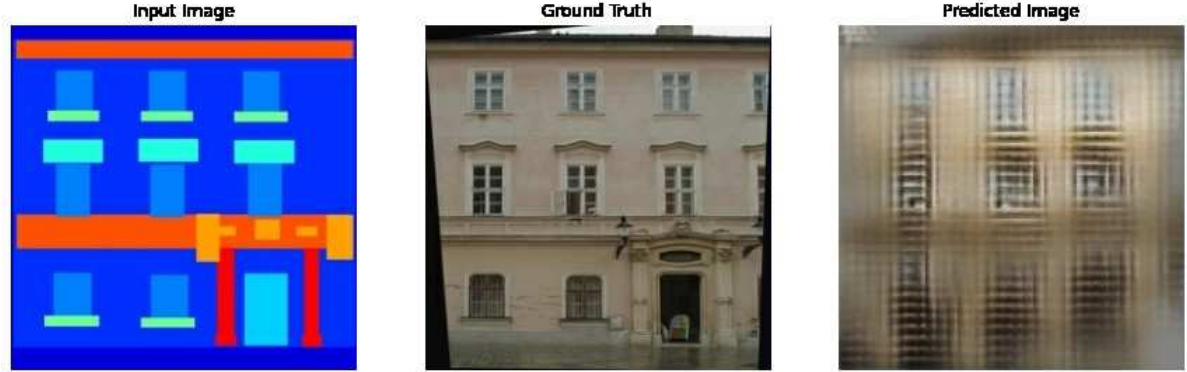}
 \end{center}
\caption{Facedes dataset results generated by custom discriminator (used transformers).}
\label{CostomDiscriminatorOutput}
\end{figure}

We also experimented with a custom PatchGAN like discriminator architecture of our creation. The proposed algorithm utilizes transformer layers instead of convolutional layers. The algorithm works by splitting the image into patches, generating patches linear embedding, adding position embeddings to preserve the positional information.  Transformer Encoders use the embedded patches as input. Finally, we used a Convolutional layer to classify $N \times N$ patches as real or fake, here $N$ is the width and height of the last convolutional layer outputs.  In this architecture, the convolutional layer stands for matching input and output shapes. Instead of a convolutional layer, up or downsampling of the transformer encoder's output is also possible. This approach does not work as well as the convolutional layer but is still usable and relatively well. The biggest drawback of not using a convolutional layer is that the generated image is pixelated and not realistic as we want them to be Figure~\ref{CostomDiscriminatorOutput} shows the result of using the experimental discriminator in the architecture.

\setlength{\parskip}{1em}

The traditional PatchGAN discriminator, made out of convolution layers, was used to obtain displayed results throughout the present paper.

\subsection{Image Quality wise Comparison with Other Architectures}

\begin{figure}[th!]
  \centering
  \begin{subfigure}[b]{\textwidth}
    \centering
    \includegraphics[width=\textwidth,height=8cm]{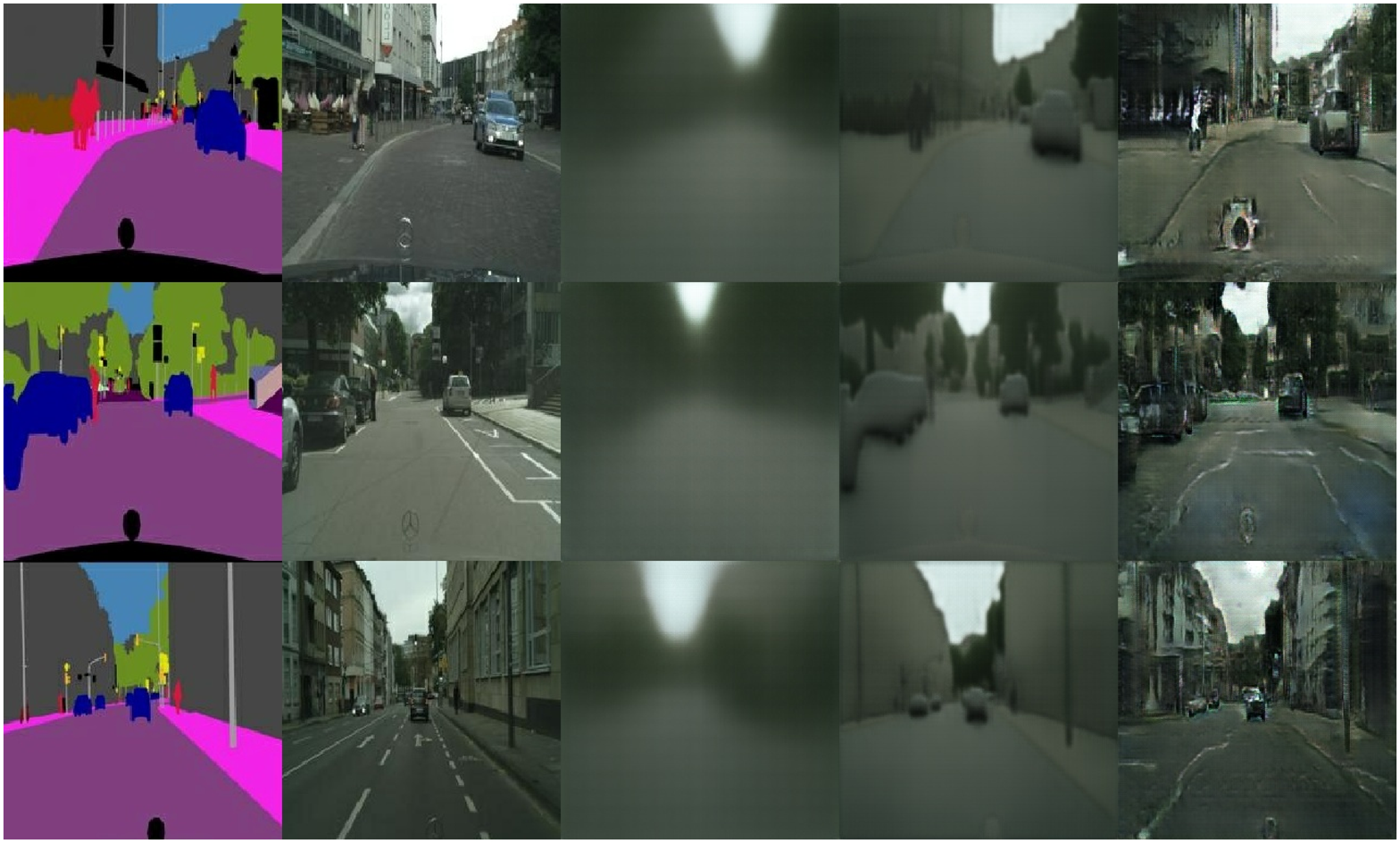}
    \vspace*{-7mm}
    \caption*{\begin{tabularx}{\textwidth}{
          | >{\centering\arraybackslash}X
          | >{\centering\arraybackslash}X
          | >{\centering\arraybackslash}X
          | >{\centering\arraybackslash}X
          | >{\centering\arraybackslash}X | }
        \hline
        Input & Target & Autoencoder & U-Net & Ours\\ \hline \end{tabularx}}
  \end{subfigure}
   \caption{Comparison of three different architectures. Photos are, left to right,  input, target, autoencoder output, U-Net output, and output obtained using our method.}
  \label{fig:ThreeModels}
\end{figure}

Evaluation of the performance of different models is not an easy task. A simple Mean Squared Error or cross-entropy is not sufficient to decide which models outperform the others. In this work, we used some of the most advanced image evaluation metrics FID(Frechet Inception Distance)~\cite{Heusel:2017}, SSIM(structural similarity)~\cite{Wang:2004}, IS(inceptions score)~\cite{Salimans:2016}. Inceptions score is one of the widely used evaluation metrics for GANs and is in most experiments. We compared our architecture with  U-Net and autoencoders that are the go-to choices for most image-to-image translation tasks such as image segmentation. We trained an autoencoder, a U-Net, and our approach on the cityscapes dataset to convert segmentation maps to photos. Table~\ref{Table:Comparisson} shows the measures of the generated image quality among the different architectures according to the metrics we mentioned above. For the visual comparison, figure~\ref{fig:ThreeModels} presents outputs of all three models.

\setlength{\parskip}{1em}

\begin{table}[h!]
\centering
 \begin{tabular}{|p{0.3\textwidth}>{\centering}p{0.2\textwidth}>{\centering}p{0.2\textwidth}>{\centering\arraybackslash}p{0.2\textwidth}|} 
 \hline
  Model & FID  & IS  & SSIM \\ [0.5ex] 
 \hline\hline
 Tensor-to-image(Ours) & 939 & 1.281 & 0.46 \\ 
 U-Net & 4318 & 1.280 & 0.36 \\
 Autoencoder & 6434 & 1.343 & 0.25 \\ [1ex] 
 \hline
 \end{tabular}
 \caption{Performance comparissons \label{Table:Comparisson}}
\end{table}



Results, shown in figure~\ref{fig:ThreeModels} and table~\ref{Table:Comparisson},  indicate that the autoencoder fails to perform as well as the competitors. It has a poor and low-level understanding of the image. Autoencoder architecture is only able to distinguish the ground from the sky, thus generates unrealistic images. The generated images mostly contain two colors, one for the sky and the other for the land. U-Net is an autoencoder with skip connections between encoder and decoder layers. U-Net was able to pick up the different objects in the scene and fill them with different colors. Despite the performance increase, the generated images were not realistic enough. They contained the contours of objects with different colors without shadows or reflections. Correct identification colors, shadows, and reflections are the main elements that make an image closer to the photo-realistic level. 

\subsection{Contribution of Adversary}

In a previous paper~\cite{Gunduc:2021} we have used a non-adversarial model which was trained with only a custom-designed generator. The results were fascinating and promising, which opened the possibility of further improvements with adversarial architecture. We have decided to extend its domain by adding a discriminator to the network and generator. We have trained both models on the facades dataset to validate the generation quality of the output. The expected is the generation of photos of the building just by the architectural labels. The models exhibit slight differences in training.  The adversarial model is trained with a cGAN alongside the L1 loss. The non-adversarial model (only generator) is optimized against the L1 loss.  The model trained with a cGAN was quite successful in generating sharp and realistic images.  The basic structures of the building, windows, doors, and edges, were in the right places.  The model optimized against the L1 loss has only a basic understanding of the shape of buildings.  Also, the generated images are blurry and not clear. So we have concluded that detailed image-to-image translation tasks require the adversary model for better performance (Figure~\ref{fig:cGANComper}).

\subsection{Object segmentation}

\begin{figure}[th!]
  \centering
  \begin{subfigure}[b]{\textwidth}
    \centering
    \includegraphics[width=\textwidth,height=8cm]{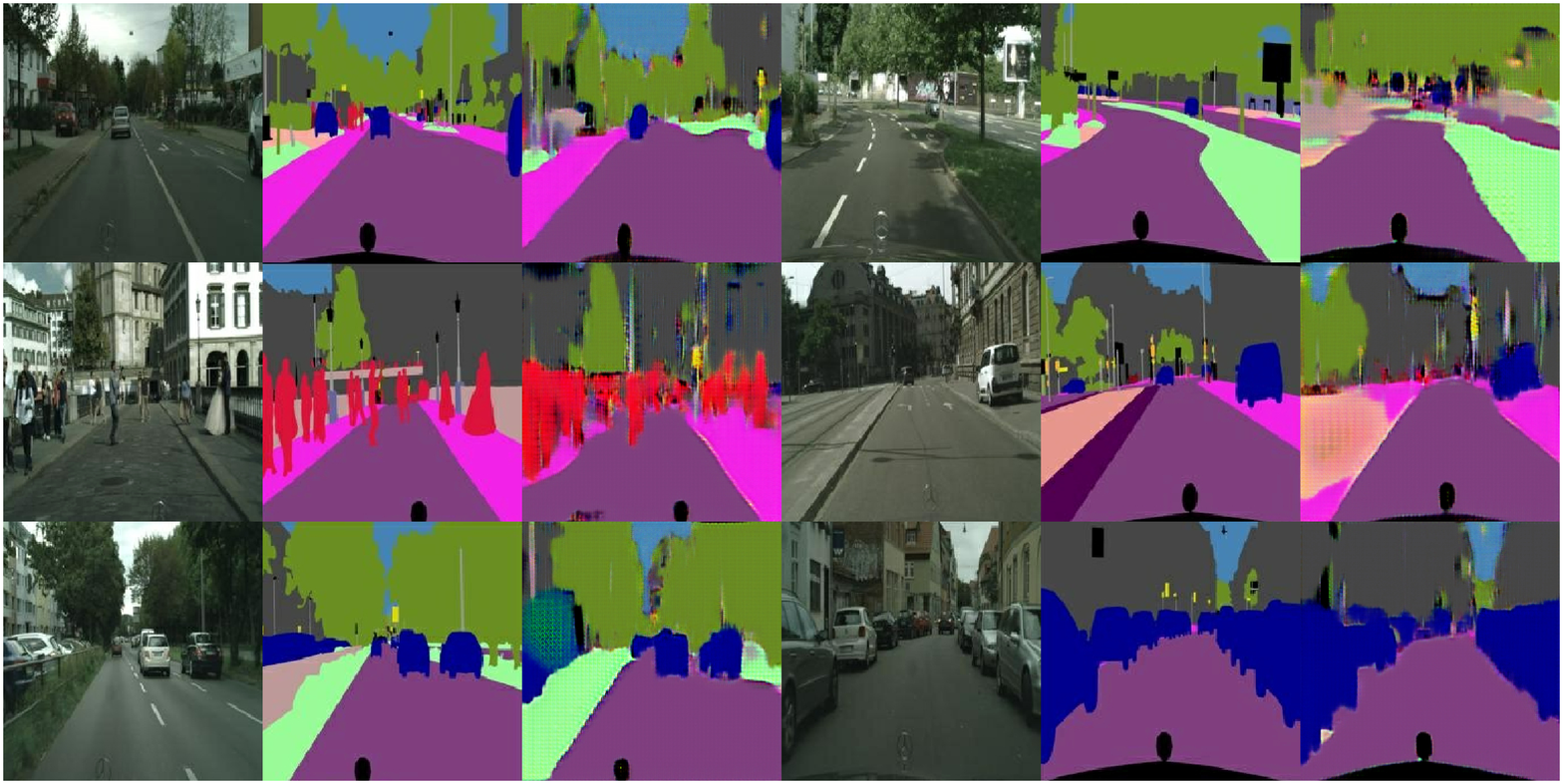}
    \vspace*{-7mm}
    \caption*{\begin{tabularx}{\textwidth}{
          | >{\centering\arraybackslash}X
          | >{\centering\arraybackslash}X
          | >{\centering\arraybackslash}X
          | >{\centering\arraybackslash}X
          | >{\centering\arraybackslash}X
          | >{\centering\arraybackslash}X | }
        \hline  
        Input & Target & Output & Input & Target & Output\\
        \hline \end{tabularx}}
  \end{subfigure}
  \caption{Generation of semantic segmentation map }
  \label{fig:Combine_s}
\end{figure}

Object segmentation is more of a vision problem than it is an image-to-image translation. It is all about finding the appropriate regions and filling them with corresponding colors. But since both the input and the output are images, we used our model to segment objects. We trained on the Cityscapes~\cite{Cordts:2016,Cordts:2015} dataset. The raw images are used as input and expected the model to output segmentation maps. It is a difficult task for a model to accomplish without a discriminator. Hence we presented  all the object segmentation results on this paper using the model trained with a discriminator (Figure~\ref{fig:Combine_s}).

\subsection{Image-to-image translation}

\begin{figure}[th!]
  \centering
  \begin{subfigure}[b]{\textwidth}
    \centering
    \includegraphics[width=\textwidth,height=8cm]{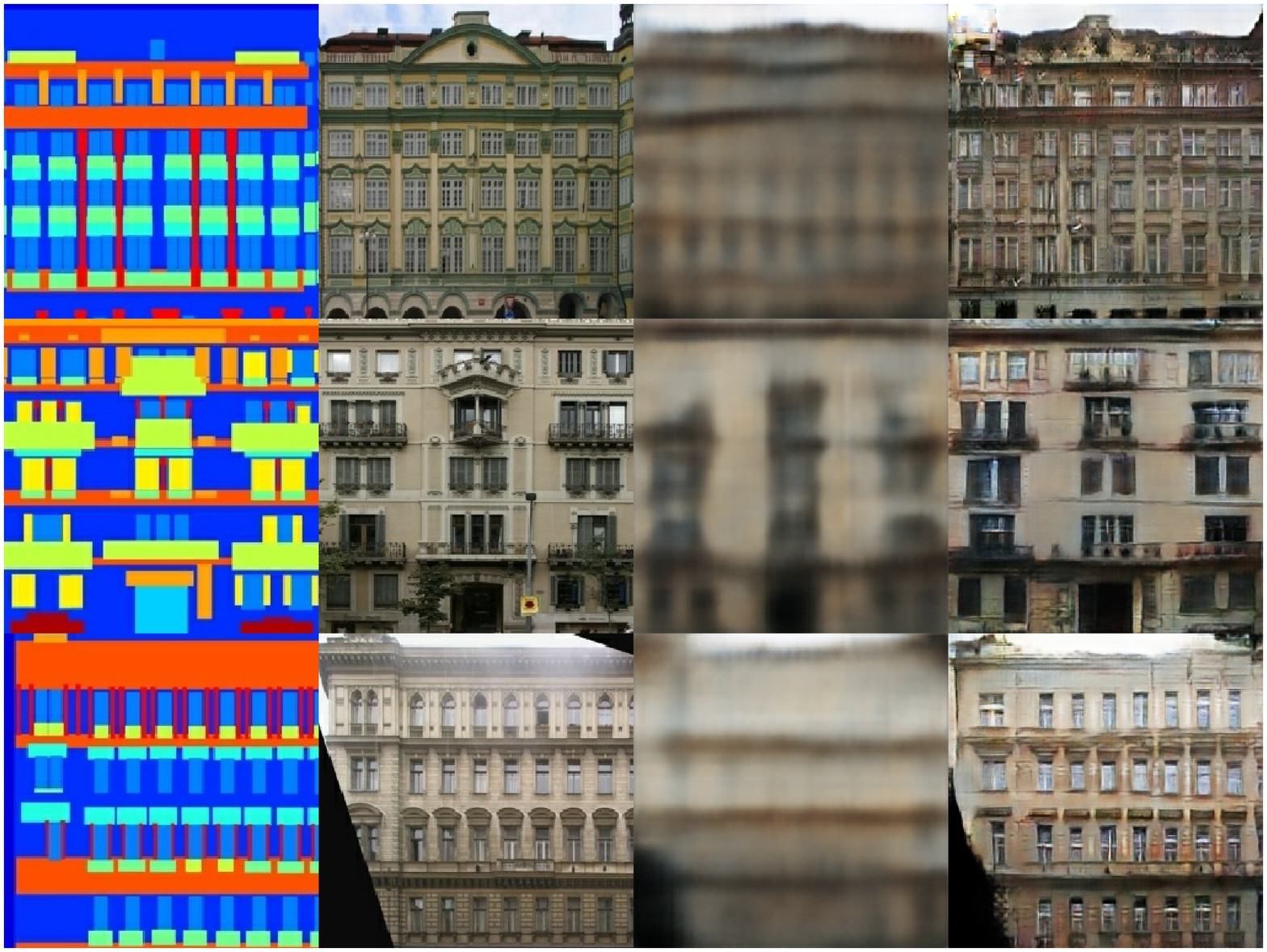}
    \vspace*{-7mm}
    \caption*{\begin{tabularx}{\textwidth}{
          | >{\centering\arraybackslash}X
          | >{\centering\arraybackslash}X
          | >{\centering\arraybackslash}X
          | >{\centering\arraybackslash}X | }
        \hline
        Input & Target & Without cGAN  & cGAN
        \\ \hline
    \end{tabularx}}
  \end{subfigure}
  \caption{Visual Comparison cGan and No cGan tets}
  \label{fig:cGANComper}
\end{figure}

Although we trained our model on different tasks, we all did them in image-to-image translation fashion where we input a 3D tensor and expected a 3D or 1D tensor. A generator-only approach is suitable for most of the tasks. Nevertheless, the model which contains a discriminator is a better solution for the close similarity between the output image and the dataset. Discriminator forces the generator to generate results that are sharp and realistic. Hence using a discriminator is the go-to choice (Figure~\ref{fig:cGANComper}).

\subsection{Depth estimation from a single image}

\begin{figure}[th!]
  \centering
  \begin{subfigure}[b]{\textwidth}
    \centering
    \includegraphics[width=\textwidth,height=8cm]{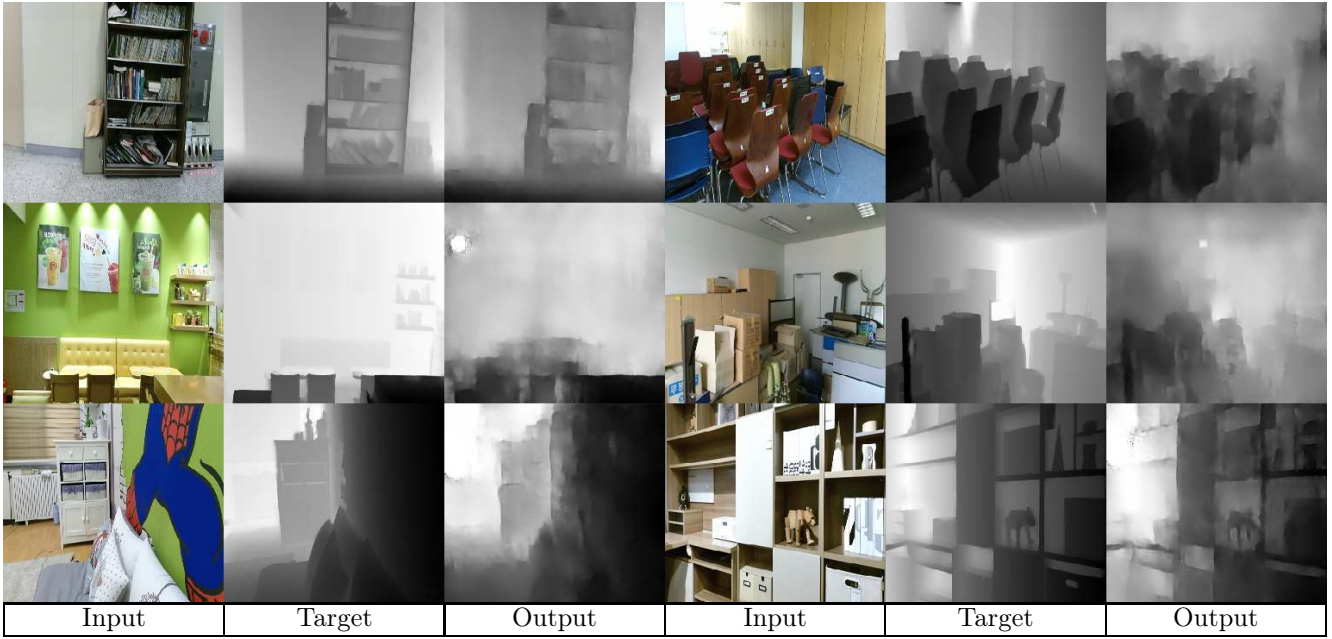}
    \vspace*{-7mm}
    \caption*{\begin{tabularx}{\textwidth}{
          | >{\centering\arraybackslash}X
          | >{\centering\arraybackslash}X
          | >{\centering\arraybackslash}X
          | >{\centering\arraybackslash}X
          | >{\centering\arraybackslash}X
          | >{\centering\arraybackslash}X | }
        \hline  
        Input & Target & Output & Input & Target & Output\\
        \hline \end{tabularx}}
  \end{subfigure}
  \caption{RGB-D Dataset: Experiment on Depth Perception of the Model }
  \label{fig:RGB-D_Depth}
\end{figure}

Estimating the depth of an image is an important problem and has a wide range of applications like self-driving cars, drones, and more. Often depth prediction is expensive because it requires specialized hardware and multiple shots of every scene. With the proposed model, we aim to generate depth maps from only one single image. This approach is economical and successful application is superion comparing with the classical two shut system.  It reduces the cost of accomplishing the depth map of an image and increases the accessibility of the depth maps. Shuts of only one camera are enough to estimate depth in our approach (Figure~\ref{fig:RGB-D_Depth}).

\section{Conclusion}

The traditional image-to-image translation methods are only applicable to a single domain or trained for predefined tasks. In this work, We used a unique vision transformers-based generator architecture to extend the range of vision tasks with a single architecture. This paper is an extension of previous work, generator-based model, Tensort-to-Image~\cite{Gunduc:2021}.  The proposed architecture uses Conditional GANs(cGANs) with a Markovian discriminator (PatchGAN). Adding a discriminator into the architecture~\cite{Gunduc:2021} improved the output image quality and the range of vision tasks. Comparing with the commonly used models, namely, autoencoder and U-Net architectures, the proposed method shows better performance on the image-to-image translation tasks.

We concluded that vision transformers are capable of performing most of the image-to-image translation tasks.  Better image quality is reinforced with an architecture containing a discriminator. The model details can be found in GitHub accont (\href{https://github.com/YigitGunduc/vit-gan}{https://github.com/YigitGunduc/vit-gan}).

\pagebreak

\pagebreak 

\pagebreak

\pagebreak

\end{document}